\documentclass[14pt]{article}
\usepackage{epsfig}
\usepackage{amsmath}
\usepackage[english]{babel}
\usepackage{amsfonts,amssymb}

\begin{document}

\renewcommand{\thesection}{\arabic{section}.}
\renewcommand{\thesubsection}{\arabic{section}.\arabic{subsection}.}
\renewcommand{\refname}{References:}
\def\fix{\varphi_\mathbf{x}}
\def\fixa{\varphi_{\mathbf{x}+\mathbf{a}}}
\def\fixaa{\varphi_{\mathbf{x}+2\mathbf{a}}}
\def\fixb{\varphi_{\mathbf{x}+\mathbf{b}}}

\title{\bf Chiral spin liquid in two-dimensional XY helimagnets}

\author{A.O.~Sorokin\footnote{aosorokin@gmail.com}, A.V.~Syromyatnikov\footnote{syromyat@thd.pnpi.spb.ru}\\
{\footnotesize\it Petersburg Nuclear Physics Institute, 188300,
Saint Petersburg, Russia }}
\date{}
\maketitle

\begin{abstract}

We carry out Monte-Carlo simulations to discuss critical
properties of a classical two-dimensional XY frustrated helimagnet
on a square lattice. We find two successive phase transitions upon
the temperature decreasing: the first one is associated with
breaking of a discrete $\mathbb{Z}_2$ symmetry and the second one
is of the Berezinskii-Kosterlitz-Thouless (BKT) type at which the
$SO(2)$ symmetry breaks. Thus, a narrow region exists on the phase
diagram between lines of the Ising and the BKT transitions that
corresponds to a chiral spin liquid.

\end{abstract}

Pacs {64.60.De, 75.30.Kz}

\maketitle

\section{Introduction}

Frustrated magnets have attracted much attention in recent years.
Exotic spin liquid phases are of special interest which have been
found in some of them \cite{Balents}. A chiral spin liquid phase
is an example of such an exotic state of matter in which there are
neither quasi long-range nor long-range magnetic orders but a
chiral order parameter $\langle {\bf S}_i\times{\bf S}_j\rangle$
is nonzero. Existence of such a phase is discussed in context of
one-dimensional frustrated quantum magnetic systems\cite{Chain},
and it is found experimentally in Ref.~\cite{ChainExp}.

In larger dimensions, one of the systems in which the chiral spin
liquid phase can be found at finite temperature is a classical
planar (XY) helimagnet with $\mathbb{Z}_2\otimes SO(2)$ symmetry
in which the helical structure results from a competition of
exchange interactions between localized spins. Critical behavior
of spin systems from this class is described by two order
parameters. Besides the conventional magnetization with $SO(2)$
symmetry, one has to take into account also the chiral order
parameter that is an Ising variable with $\mathbb{Z}_2$ symmetry.
This parameter characterizes the direction of the helix twist and
distinguishes left-handed and right-handed helical structures.

In three-dimensional helimagnets, the phase transitions on the
magnetic and the chiral order parameters occur simultaneously. It
was found numerically  that the transition is of the weak first
order or of the "almost second order"\cite{Zumbach, DMT} type in
helical antiferromagnets on a body-centered tetragonal lattice
\cite{Helix} and on a simple cubic lattice with an extra competing
exchange coupling along one axis \cite{Sorokin}. These systems
belong to the same (pseudo)universality class as, e.g., the model
on a stacked-triangular lattice \cite{Kaw1} and $V_{2,2}$ Stiefel
model \cite{KunZumb}. The possibility of existence and
stabilization of the chiral spin liquid phase by, e.g.,
Dzyaloshinsky-Moria interaction in 3D helimagnets is discussed
recently in Refs.~\cite{Onoda}.

In two dimensions, the situation is rather different
\cite{Olsson1}. Two successive transitions were observed with the
temperature decreasing. The chiral order appears as a result of
the first transition that is of the Ising type. Another one is the
Berezinskii-Kosterlitz-Thouless (BKT) transition driven by the
unbinding of vortex-antivortex pairs \cite{BKT}. Then, the chiral
spin liquid phase arises between these transitions with the chiral
order and without a magnetic one. Various 2D systems from the
class $\mathbb{Z}_2\otimes SO(2)$ were investigated numerically
(see Ref. \cite{Hasenbush} for review): triangular antiferromagnet
\cite{Triangular, TriangNNN}, J$_1$-J$_2$ model \cite{JJ}, the
Coulomb gas system of half-integer charges \cite{Gas}, two coupled
XY models \cite{XY-XY}, Ising-XY model \cite{Ising-XY, Hasenbush,
CFT} and the generalized fully frustrated XY model \cite{GFFXY}.
And surely, the most famous of them is the fully frustrated XY
model (FFXY) introduced by Villain \cite{Villain}. This model is
of great interest because it describes a superconducting array of
Josephson junctions under an external transverse magnetic field
\cite{Teitel}. It was found that the temperature of the Ising
transition $T_I$ is $1\mbox{-}3\%$ larger than that of the BKT
transition for the most of above-named systems \cite{Teitel,
Olsson1, FFXY, Hasenbush}.

Korshunov argued \cite{Korshunov} that a phase transition, driven
by unbinding of kink-antikink pairs on the domain walls associated
with the $\mathbb{Z}_2$ symmetry, can take place in models similar
to 2D FFXY one at temperatures appreciably smaller than $T_{BKT}$
(see also Ref.~\cite{Kinks}). Such a transition could lead to a
decoupling of phase coherence across domain boundaries, producing
in this way two separate bulk transitions with $T_{BKT}<T_{I}$
\cite{Tw}. It was pointed out however in Ref.~\cite{Korshunov}
that these two continuous transitions can merge into a single
first order one. These conclusions do not depend on the particular
form of interactions in system as soon as the ground state
degeneracy remains the same. They are confirmed by numerical
studies of the models mentioned above \cite{Teitel, Olsson1, FFXY,
Hasenbush, JJ, Triangular, TriangNNN, Gas}.

Nevertheless the situation remains contradictory in 2D helimagnets
belonging to the same $\mathbb{Z}_2\otimes SO(2)$ class as FFXY
model and the antiferromagnet on the triangular lattice. Garel and
Doniach \cite{Garel} (see also \cite{Okwamoto}) considered the
simplest helimagnet on a square lattice with an extra competing
exchange coupling along one axis that is describing by the
Hamiltonian
\begin{equation}
    H=\sum\limits_{\mathbf{x}}\Bigl(J_1\cos(\fix-\fixa)
                                 +J_2\cos(\fix-\fixaa)
                                 -J_b\cos(\fix-\fixb)\Bigr),
    \label{ham}
\end{equation}
where the sum runs over sites $\mathbf{x}=(x_a,x_b)$ of the
lattice, $\mathbf{a}=(1,0)$ and $\mathbf{b}=(0,1)$ are unit
vectors of the lattice, the coupling constants $J_{1,2}$ are
positive. Using arguments of Ref.~\cite{Einhorn}, they concluded
\cite{Garel} that at low temperatures the vertices are bound by
strings, which would inhibit the BKT transition and make the Ising
transition occur first with the temperature increasing. Kolezhuk
noticed \cite{Kolezhuk} that those arguments are not valid for a
helimagnet, and showed that the Ising transition temperature is
larger than the BKT one at least near the Lifshitz point
$J_2=J_1/4$. It was found by Monte Carlo simulations in the recent
paper \cite{Cinti} that $T_{BKT}>T_{I}$ at $J_2=0.3$ and
$J_1=J_b=1$ (i.e., very near the Lifshitz point) in accordance
with Ref.~\cite{Garel} and in contrast to Ref.~\cite{Kolezhuk}.

\begin{figure}
    \center
    \vspace{-7mm}
    \includegraphics[height=80mm]{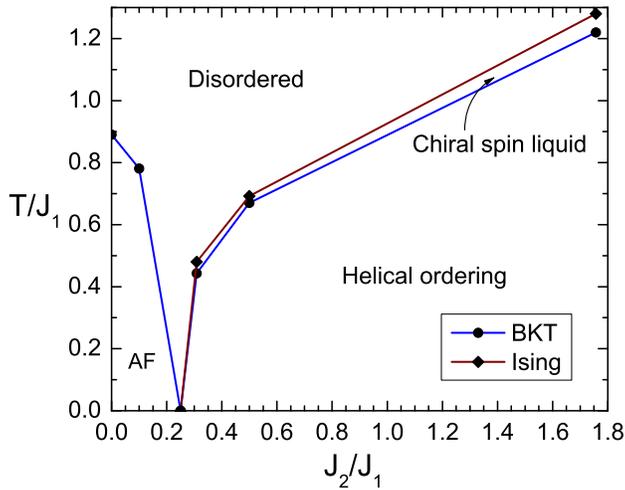}
    \vspace{-7mm}
    \caption{{\small Phase diagram of the model \eqref{ham} that is found in the present paper.}}
    \label{phase}
\end{figure}

To account for the discordance between results for helical magnets
and the general arguments for $\mathbb{Z}_2\otimes SO(2)$ class,
we perform extensive Monte Carlo simulations of the model (1) for
different values of $J_2$. We obtain reliable results at
$J_2>0.4J_1$ which show that $T_{BKT}<T_I$. On the other hand the
value of $T_I$ close to the Lifshitz point is hiding among effects
of the finite size scaling and is not accessible for ordinary
estimation methods. We obtain the Ising transition temperature
from the chiral order parameter distribution and find that
$T_{BKT}<T_I$ near the Lifshitz point too. At the same time we
find in accordance with results of Ref.~\cite{Cinti} that the
specific heat and susceptibilities have subsidiary peaks at low
$T<T_{BKT}$ near the Lifshitz point. These are anomalies which are
attributed in Ref.~\cite{Cinti} to the Ising phase transition.
However, we demonstrate that these anomalies do not signify a
continuous phase transition. Apparently, their origin is in
metastable states which lead also to a peculiar distribution of
the chiral order parameter. We find no such features in the
specific heat and susceptibilities far from the Lifshitz point (at
$J_2>0.4J_1$). As a result we obtain the phase diagram shown in
Fig.~\ref{phase}.

The rest of the present paper is organized as follows. We discuss
in Sec.~\ref{model} the model \eqref{ham} in more detail and
introduce quantities to be found in our calculations. Numerical
results are discussed in Sec.~\ref{res}. In particular, the Ising
and the BKT transitions are considered in Secs.~\ref{ising} and
\ref{bkt}, respectively. The neighborhood of the Lifshitz point
and the phase diagram are discussed in Sec.~\ref{phsec}.
Sec.~\ref{conc} contains our conclusions.

\section{Model and methods}
\label{model}

We consider the model (\ref{ham}) of the classical XY magnet on a
square lattice. We set $J_1=J_b=1$ for simplicity and the value of
the extra exchange interaction $J_2$ is a variable. The Lifshitz
point corresponds to $J_2=1/4$ in this notation. The system has a
collinear antiferromagnetic ground state at $J_2<1/4$. To discuss
the phase transition from the (quasi-)antiferromagnetic phase to
the paramagnetic one we consider $J_2=0$ and $J_2=0.1$ (see
Fig.~\ref{phase}). The ground state has a helical ordering at
$J_2>1/4$. The turn angle $\theta_0$ between two neighboring spins
along $\mathbf{a}$ axis is given by $\cos\theta_0=-J_1/4J_2$ at
zero temperature.

To discuss the number and the sequence of phase transitions from
the (quasi-)helical phase to the paramagnetic one we consider
$J_2\approx 0.309$, 0.5 and $1.76$ corresponding at $T=0$ to
angles of commensurate helices $\theta_0=4\pi/5$, $2\pi/3$ and
$6\pi/11$, respectively. We use lattices with $L^2$ cites, where
$L$ is divisible by the size of the helix pitch and it lies in the
range from 20 to 120. We apply the periodic (toric) boundary
conditions as well as the cylindrical ones (i.e., with the
periodic condition along the $\mathbf{b}$ axis and the free one
along the $\mathbf{a}$ axis). We have found that both conditions
lead to the same values of transition temperatures and indexes. In
contrast values of Binder's cumulants and the chiral order
parameter distribution at $J_2\approx 0.309$ depend on boundary
conditions as we discuss below in detail. Standard Metropolis
algorithm \cite{Metropolis} has been used. The thermalization was
maintained within $4\cdot10^5$ Monte Carlo steps in each
simulation. Averages have been calculated within $3.6\cdot10^6$
steps for ordinary points and $6\cdot10^6$ for points close to the
critical ones. We have used also the histogram analysis technique
in which the range of each quantity has been divided into
$6.4\cdot10^5$ bins.

\subsection{Order parameters}

The BKT transition is driven by the magnetic order parameter for
which we use two definitions. Similar to the triangular lattice
\cite{Triangular}, one can introduce a number of sublattices in
the case of helix pitches which are divisible by the lattice
constant. Then, one can write for the magnetic order parameter
\begin{equation}
    \mathbf{m}_i=\frac{n_{sl}}{L^2}\sum_{\mathbf{x}_i}\mathbf{S}_{\mathbf{x}_i},\quad
    \overline{m}=\sqrt{\frac1{n_{sl}}\sum\limits_{i}\left<\mathbf{m}_i^2\right>},
    \label{m}
\end{equation}
where index $i$ enumerates $n_{sl}$ sublattices, the sum over
$\mathbf{x}_i$ runs over sites of the $i$-th sublattice, spin
$\mathbf{S}_{\mathbf{x}_i}=(\cos\phi_{\mathbf{x}_i},\sin\phi_{\mathbf{x}_i})$
is a classical two-component unit vector, and
$\langle\ldots\rangle$ denotes the thermal average. The second
definition of the order parameter is valid both for commensurate
and incommensurate helices
\begin{equation}
    \mathbf{M}_j=\frac1{L}\sum_i\mathbf{S}_{j\mathbf{a}+i\mathbf{b}},\quad
    \overline{M}=\sqrt{\frac1{L}\sum_{j}\left<\mathbf{M}_j^2\right>}.
    \label{ml}
\end{equation}
Our calculations show that definitions \eqref{m} and \eqref{ml}
lead to the same results away from the Lifshitz point. We have
found that $\bf m$ shows an anomalous behavior at $J_2\approx1/4$
and we use definition \eqref{ml} in this case. Thus, we
demonstrate that it is useful in numerical discussion of
helimagnets to choose parameters of the Hamiltonian so that the
helix pitch at $T=0$ to be commensurate.

The Ising transition is driven by the
chiral order parameter defined as
\begin{equation}
    k=\frac{1}{L^2\sin\theta_0}\sum\limits_\mathbf{x} \sin(\fix-\fixa),
    \quad \overline{k}=\sqrt{\left<k^2\right>}.
    \label{kiral}
\end{equation}

\subsection{Susceptibilities and cumulants}

We introduce corresponding susceptibilities for all order
parameters \cite{Binder}
\begin{equation}
    \chi_p=\left\{
    \begin{array}{ll}
    \displaystyle \frac{L^2}{T}\left(\left<p^2\right>-\left<|p|\right>^2\right), \quad & T<T_c, \\
        &\\
        \displaystyle \frac{L^2}{T}\left<p^2\right>, \quad & T\ge
        T_c.
    \end{array}
    \right.
    \label{suscept}
\end{equation}
The second line in this definition is used below for estimation of critical
exponents. Binder's cumulants \cite{Binder} are define as
\begin{equation}
    U_p=1-\frac{\langle p^4 \rangle}{3\langle p^2 \rangle^2}.
        \label{up}
\end{equation}
We discuss also the cumulant
\begin{equation}
V_k=\frac{\partial}{\partial(1/T)}\ln
\langle k^2\rangle=L^2\left(\frac{\left<k^2E\right>}{\left<k^2\right>}-\langle E\rangle\right),
\label{Vp}
\end{equation}
using which the critical exponent $\nu_k$ can be found by
finite-size scaling analysis \cite{Ferren}.

\subsection{Helicity modulus}
\label{helmod}

It is useful to introduce the helicity modulus (or the spin
stiffness) \cite{Helicity} to discuss the BKT transition that is
defined by the increase in the free energy density $F$ due to a
small twist $\Delta_\mu$ across the system in one direction ($\bf
a$ or $\bf b$)
\begin{equation}
    \Upsilon_{\mu}=\left.\frac{\partial^2 F}{\partial{\Delta_\mu}^2}\right|_{\Delta_\mu=0},
    \label{Y=dF po dd}
\end{equation}
where $\mu=a,b$ denotes the direction. Important universal
properties of a BKT transition predicted by Kosterlitz and Nelson
\cite{Jump} are the jump of the helicity modulus (\ref{Y=dF po
dd}) from zero at $T>T_{BKT}$ to the value of $2T_{BKT}/\pi$ at
$T=T_{BKT}$ and the value of the exponent $\eta(T=T_{BKT})=1/4$.
These properties have become standard methods of finding the
transition temperature.

As a result of the fact that the exchange couplings along
$\mathbf{a}$ and $\mathbf{b}$ axes are different, the helicity
moduli in these directions differ too. Thus, at zero temperature
$\Upsilon_a(0)=4J_2-J_1^2/(4J_2)$, while $\Upsilon_b(0)=J_b$.
Nevertheless, both $\Upsilon_a$ and $\Upsilon_b$ must vanish at
the same temperature with the identical value of the jump.

One finds after trivial calculations using Eqs.~\eqref{ham} and
\eqref{Y=dF po dd} that the helicity modulus $\Upsilon_b$ is
expressed via correlation functions and has a common view
\cite{Ohta}
\begin{equation}
    \Upsilon_b = \left<E''_b\right>-\frac{L^2}{T}\left<(E'_b)^2\right>,
    \label{Yy}
\end{equation}
where $T$ is measured in units of $J_1=J_b=1$, we set $k_B=1$,
$E'_b = L^{-2}\sum_\mathbf{x}\sin(\fix-\fixb)$ and $E''_b =
L^{-2}\sum_\mathbf{x}\cos(\fix-\fixb)$. Similar calculations give
for the helicity modulus in the $\mathbf{a}$ direction
\begin{equation}
    \Upsilon_a=\left<E''_a\right>-\frac{L^2}{T}\left<(E'_a)^2\right>+
    \frac{L^2}{T}\left<E'_a\right>^2,
    \label{Yx}
\end{equation}
where $E'_a = L^{-2}\sum_\mathbf{x}
(\sin(\fix-\fixa)+2J_2\sin(\fix-\fixaa) )$ and $E''_a =
-L^{-2}\sum_\mathbf{x}(\cos(\fix-\fixa)+4J_2\cos(\fix-\fixaa))$.
It may seem that the last term in Eq.~\eqref{Yx} can be discarded
as it is done with the corresponding term in Eq.~\eqref{Yy} which
is equal to zero. However it is not so because
$\left<E'_b\right>=0$ at all $T$ whereas $\left<E'_a\right>=0$ at
$T\ge T_I$ and $\left<E'_a\right>\ne0$ at $T<T_I$. To demonstrate
this let us apply an infinitesimal twist $\Delta_a$ across the
system in the $\bf a$ direction, i.e., let us replace in
Eq.~\eqref{ham} $\fix-\fixa$ by $\fix-\fixa+\Delta_a$. One writes
in the first order in $\Delta_a$
\begin{eqnarray}
&&\sum_{\bf x}\Bigl(\cos(\fix-\fixa+\Delta_a)+J_2\cos(\fix-\fixaa+2\Delta_a)\Bigr)\nonumber\\
&&\approx
\sum_{\bf x}\Bigl(\cos(\fix-\fixa)+J_2\cos(\fix-\fixaa)\Bigr)
\nonumber\\ &&-\Delta_a \sum_{\bf
x}\Bigl(\sin(\fix-\fixa)+2J_2\sin(\fix-\fixaa)\Bigr).
\label{exp}
\end{eqnarray}
Comparing the last term in Eq.~\eqref{exp} with the chiral order
parameter $k$ definition \eqref{kiral} and noting that one can use
an equivalent definition $\tilde k=L^{-2}\sum_\mathbf{x}
\sin(\fix-\fixaa)$ we conclude that the last term in
Eq.~\eqref{exp} is a linear combination of $k$ and $\tilde k$.
However, $k$ and $\tilde k$ have opposite signs in the case
considered. In particular, their combination $k+2J_2\tilde k$ in
the last term in Eq.~\eqref{exp} is equal to zero at $T=0$.
Nevertheless our numerical results presented below show that this
combination is not equal to zero at $T\ne0$ and it can be
considered as the Ising order parameter at $T\sim T_I$. Then, one
see from Eq.~\eqref{exp} that $\Delta_a$ plays the role of the
"chiral" field and, consequently, $\partial
F/\partial\Delta_a|_{\Delta_a=0}$ (that is equal in our notation
to $\left<E'_a\right>$) is proportional to the chiral order
parameter which is equal to zero at $T\ge T_I$ and which is finite
at $T<T_I$.

\begin{figure}[t]
    \vspace{-7mm}
    \parbox{0.48\textwidth}{
    \centering
    \includegraphics[height=70mm]{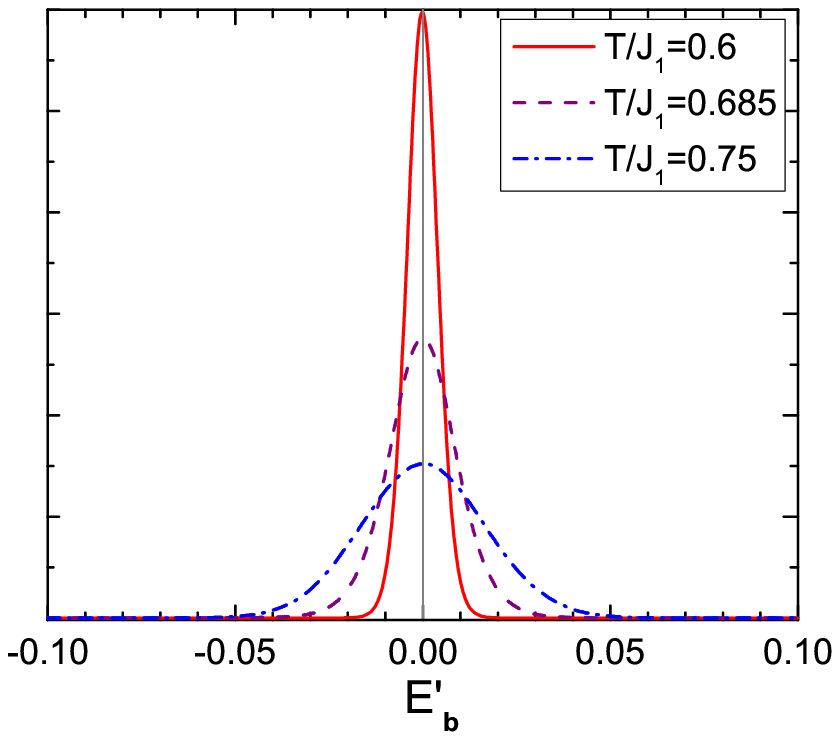}
    \vspace{-7mm}
    \caption{{\small Distribution of the value $E_b'$ defined in Eq.~\eqref{Yy} for $J_2=0.5$, $L=$ and three $T$ values: $T>T_I$, $T<T_{BKT}$, and $T_{BKT}<T<T_I$.}}
    \label{fig11}}
    \hspace{0.02\textwidth}
    \parbox{0.48\textwidth}{
    \center
    \vspace{-4mm}
    \includegraphics[height=70mm]{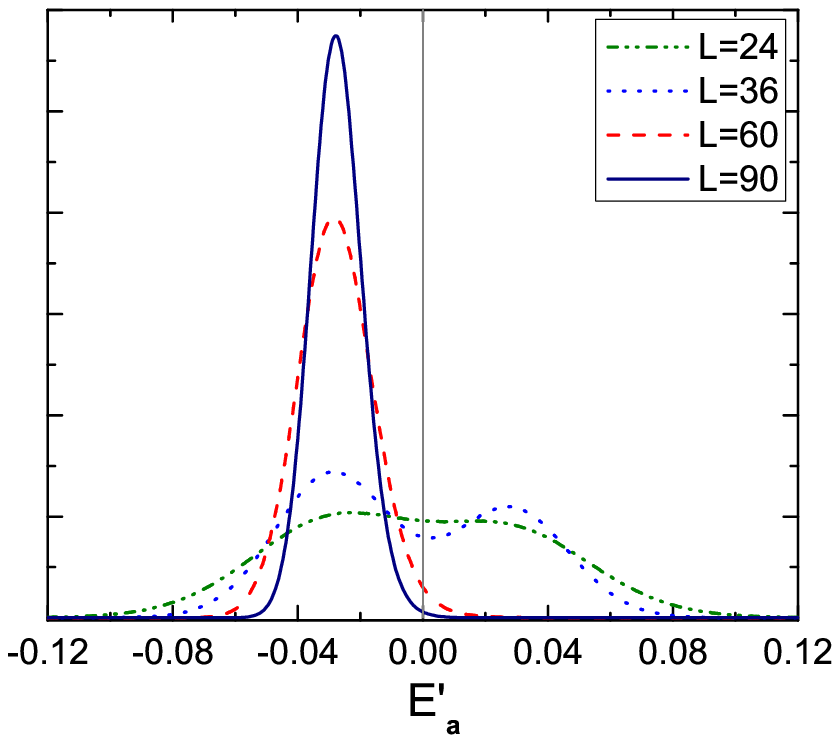}
    \vspace{-7mm}
    \caption{{\small Distribution of the value $E_a'$ defined in Eq.~\eqref{Yx} for $J_2=0.5$, $T=0.67<T_I$ and different $L$.}}
    \label{fig12}}
\end{figure}

\begin{figure}[t]
    \vspace{-7mm}
    \centering
    \includegraphics[height=70mm]{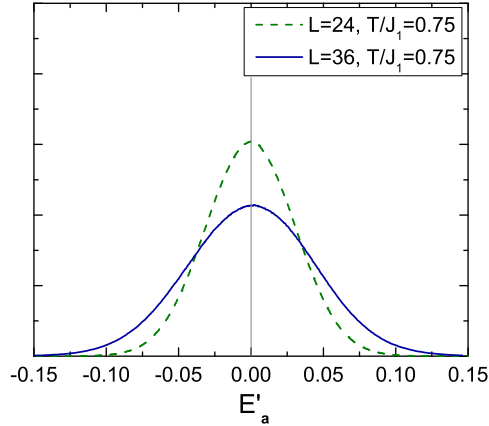}
    \vspace{-7mm}
    \caption{{\small Same as in Fig.~\ref{fig12} for $T>T_I$.}}
    \label{fig13}
\end{figure}

To illustrate this consideration we present in Fig.~\ref{fig11}
the distribution of $E'_b$ for $J_2=0.5$ and for various
temperatures below $T_{BKT}\approx0.671$, between $T_{BKT}$ and
$T_{I}\approx0.69$, and above $T_I$ (the values of $T_I$ and
$T_{BKT}$ are obtained below). The distribution has a Gaussian
form with the zero expected value $\langle E'_b\rangle=0$.
Fig.~\ref{fig12} shows the distribution of $E'_a$ for $T=0.67<T_I$
and various lattice sizes. A non-zero expected value of $\langle
E'_a\rangle$ is seen. One can observe a double-peak structure of
$E'_a$ distribution (see Fig.~\ref{fig12}) for small lattices or
at $T$ that is close enough to $T_I$, when the system can tunnel
to a configuration with opposite chirality. The probability of
such tunneling is estimated as
\begin{equation}
   p(\overline{k}_+\to\overline{k}_-)=\exp\left(-\frac{2Lf_{dw}}{T}\right),
\end{equation}
where $f_{dw}$ is a domain wall tension \cite{domainwall} that is
positive at $T<T_I$ and it vanishes at $T=T_I$. That is why we
observe two peaks for small lattices and only one peak for large
ones. Quite expectedly, we find the single-peak distribution of
$E'_a$ at $T>T_I$ demonstrated in Fig.~\ref{fig13}.

It should be stressed that the disappearance of the double-peak
structure at the critical temperature $T_I(L)$ is a signature of
the transition on a lattice with size $L$. The value of $T_I(L)$
is close to the correct value of the transition temperature for
large lattices. We use this circumstance below in our analysis of
the Lifshitz point neighborhood.

It should be noted also that we replace in our numerical
calculations $\left<E'_a\right>$ by $\left<|E'_a|\right>$ at
$T<T_I$ in the last term in Eq.~\eqref{Yx} as it is usually done
in considerations of order parameters \cite{Binder}. It is done
because the order parameter distribution has tails in both
positive and negative regions even below the transition
temperature. The value $\left<p\right>$ is replaced by
$\left<|p|\right>$ in Eq.~\eqref{suscept} by the same reason.

\section{Numerical results}
\label{res}

We discuss in this section in detail our results for the special
case of $J_2=0.5$ that corresponds at $T=0$ to 120$^\circ$ helical
structure with three sublattices. Then, we discuss the phase
diagram. We consider lattices with $L=24$, 30, 36, 42, 48, 60, 72,
90, 120. The lattice with $L=18$ is also used for estimation of
some quantities.

\subsection{Ising transition}
\label{ising}

\begin{figure}[t]
    \vspace{-7mm}
    \parbox{0.48\textwidth}{
    \centering
    \includegraphics[height=65mm]{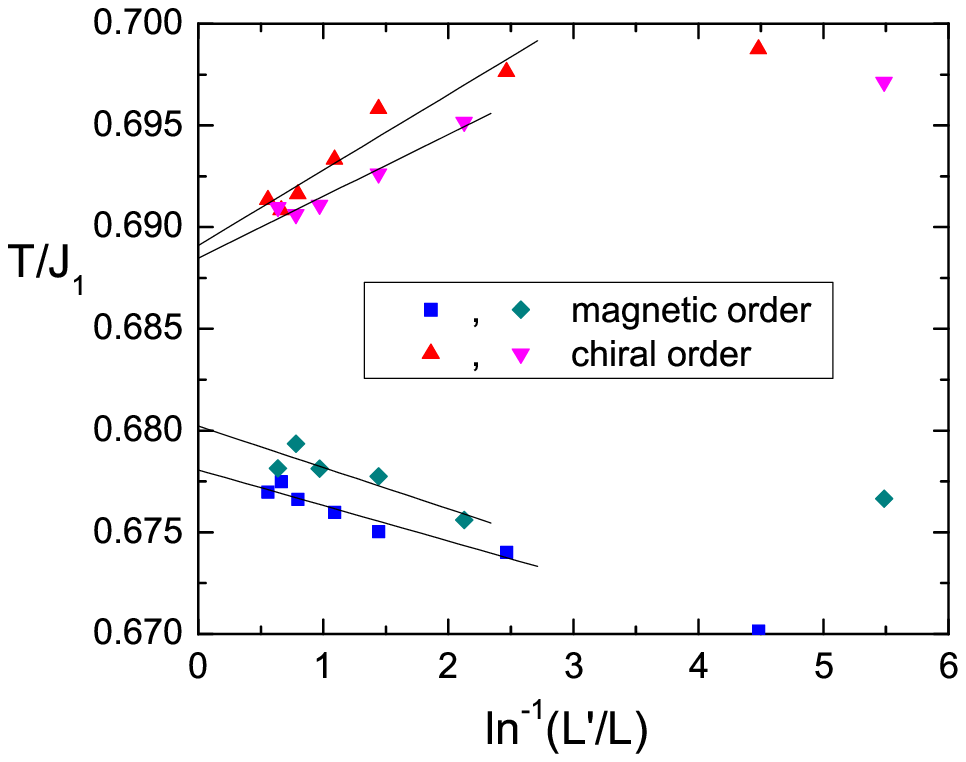}
    \vspace{-9mm}
    \caption{{\small Estimation of the transitions temperature by
    the Binder cumulant crossing method.}}
    \label{fig1}}
    \hspace{0.02\textwidth}
    \parbox{0.48\textwidth}{
    \center
    \includegraphics[height=65mm]{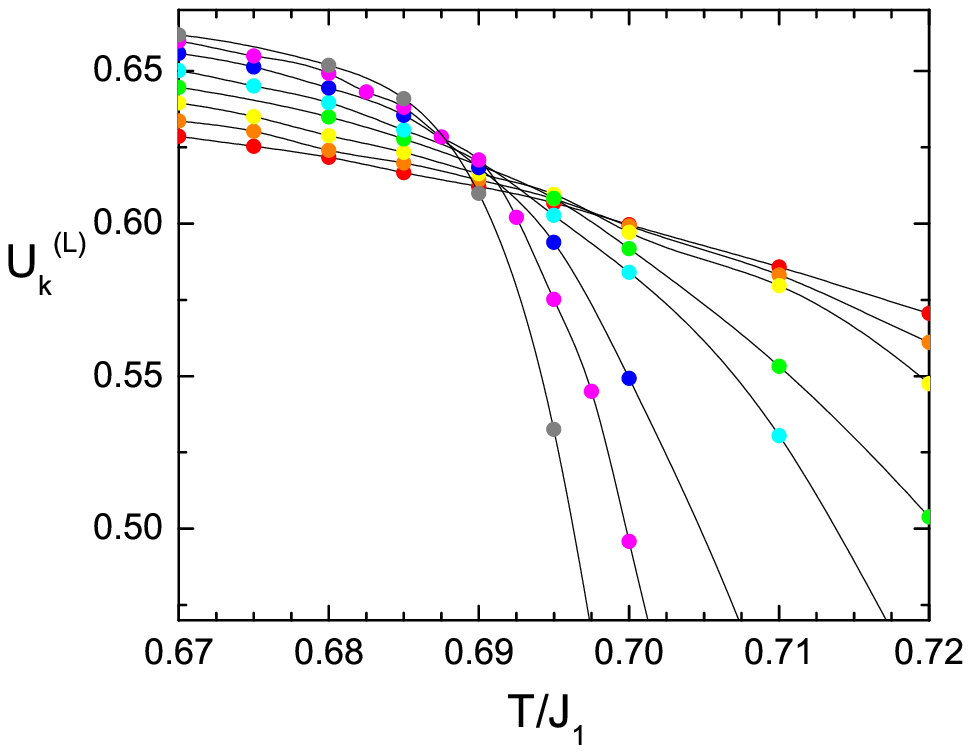}
    \vspace{-8mm}
    \caption{{\small Binder's cumulant $U_k(L)$ defined by Eq.~\eqref{up} as a function of
    temperature, for $L=24,\ldots,90$.}}
    \label{fig2}}
\end{figure}

To obtain the Ising transition temperature $T_I$ we use the Binder
cumulant crossing method \cite{Binder}. We find for $L=24$ and 30
the temperature $T_{L'}$ as a function of
$\ln^{-1}\left(L'/L\right)$ at which curves $U_k(L)$ intersect for
different lattice sizes $L'>L$. Extrapolation to the thermodynamic
limit $L'\to\infty$ gives the following transition temperature
(see Figs.~\ref{fig1} and \ref{fig2})
\begin{equation}
\label{ti}
    T_I=0.689(1).
\end{equation}
The dispersion in the value of $T_I$ obtained for the different
lattice sizes $L$ gives the error of the transition temperature
estimation. Notice that error bars are not shown in
Figs.~\ref{fig1} and \ref{fig2} and in all figures below if they
are smaller than or comparable with symbols size.

\begin{figure}[h]
    \vspace{-7mm}
    \parbox{0.48\textwidth}{
    \centering
    \includegraphics[height=65mm]{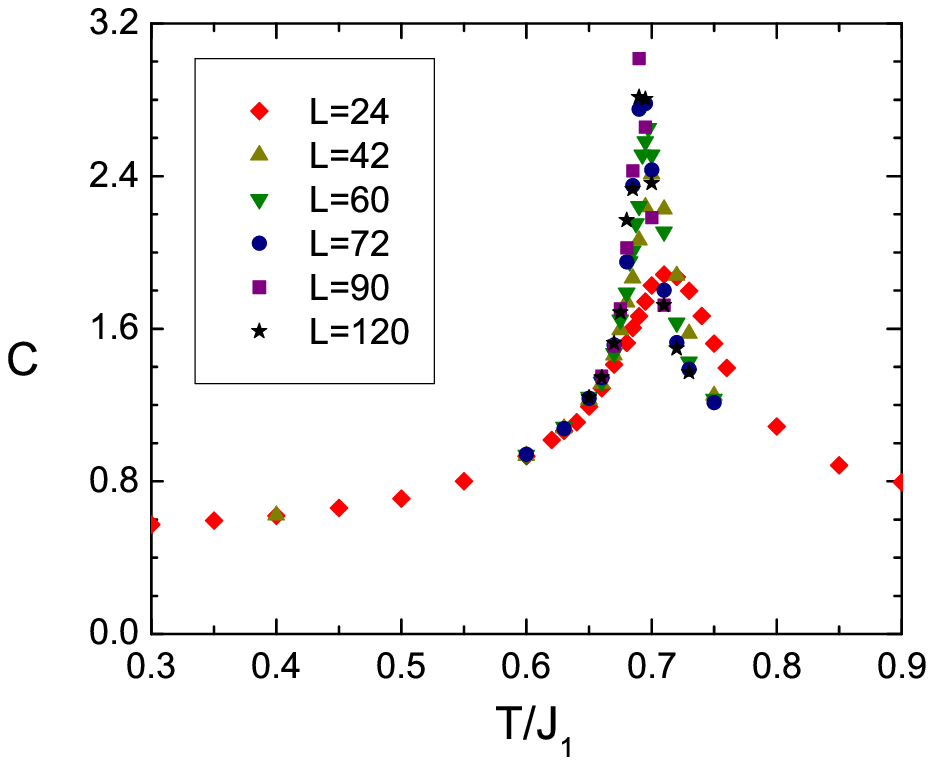}
    \vspace{-7mm}
    \caption{{\small Specific heat $C(T)$.}}
    \label{fig3}}
    \hspace{0.02\textwidth}
    \parbox{0.48\textwidth}{
    \center
    \includegraphics[height=65mm]{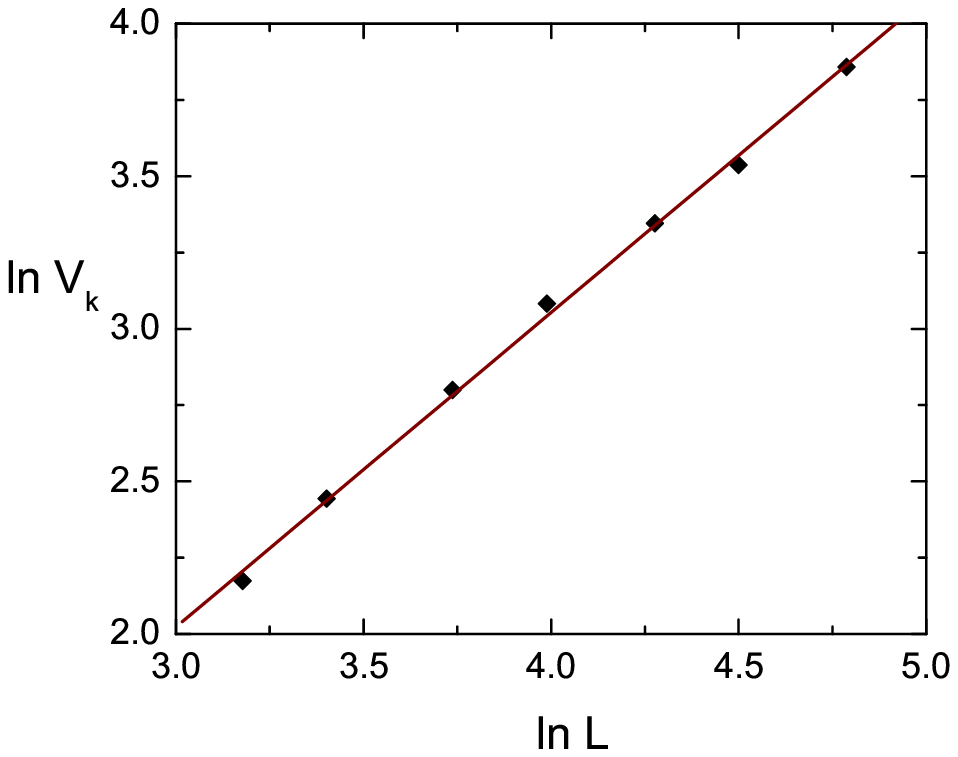}
    \vspace{-7mm}
    \caption{{\small Estimation of the exponent $\nu_k$ using the cumulant $V_k$ defined by Eq.~\eqref{Vp}.}}
    \label{fig4}}
\end{figure}

A peak in the specific heat shown in Fig.~\ref{fig3} is found
approximately at the same temperature. For the largest lattices
the peak is located in the range of temperature from $0.690$ to
$0.695$. One expects that this peak corresponds to the logarithmic
divergence of the specific heat that is a characteristic of the 2D
Ising model in which the critical exponent $\alpha$ is equal to
zero. Insufficient accuracy of our data for the specific heat
prevents us from the immediate estimation of $\alpha$.

Critical exponents $\nu_k$, $\beta_k$ and $\gamma_k$ are obtained
by the finite-size scaling theory. To estimate the exponent $\nu_k$,
we find a maximum of the quantity $V_k$ given by Eq.~(\ref{Vp}) as a
function of lattice size $L$ \cite{Ferren}
\begin{equation}
    \left(V_k^{(L)}\right)_{\mbox{\scriptsize max}}\sim
    L^{1/\nu_k}.
\end{equation}
The fitting presented in Fig.~\ref{fig4} gives
\begin{equation}
    \nu_k=0.97(4)
\end{equation}
that coincides within the computational error with the exact value of
$\nu=1$ for the 2D Ising model.

\begin{figure}[t]
    \vspace{-7mm}
    \parbox{0.48\textwidth}{
    \centering
    \includegraphics[height=65mm]{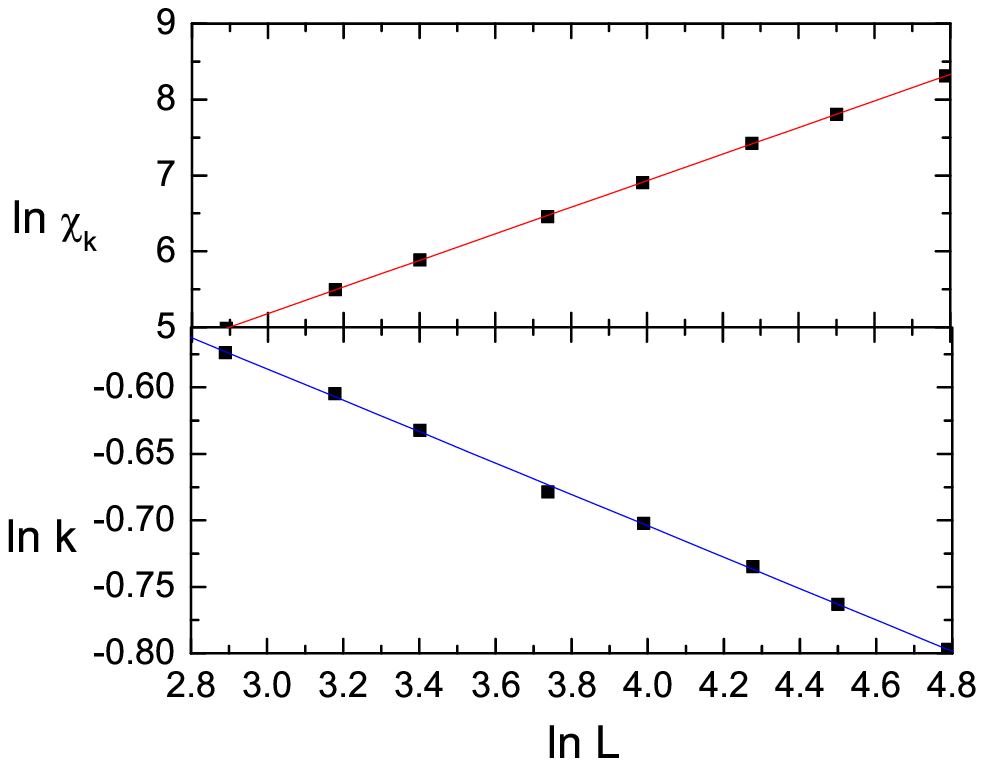}
    \vspace{-7mm}
    \caption{{\small Estimation of the exponents $\beta_k$ and $\gamma_k$.}}
    \label{fig5}}
    \hspace{0.02\textwidth}
    \parbox{0.48\textwidth}{
    \center
    \includegraphics[height=65mm]{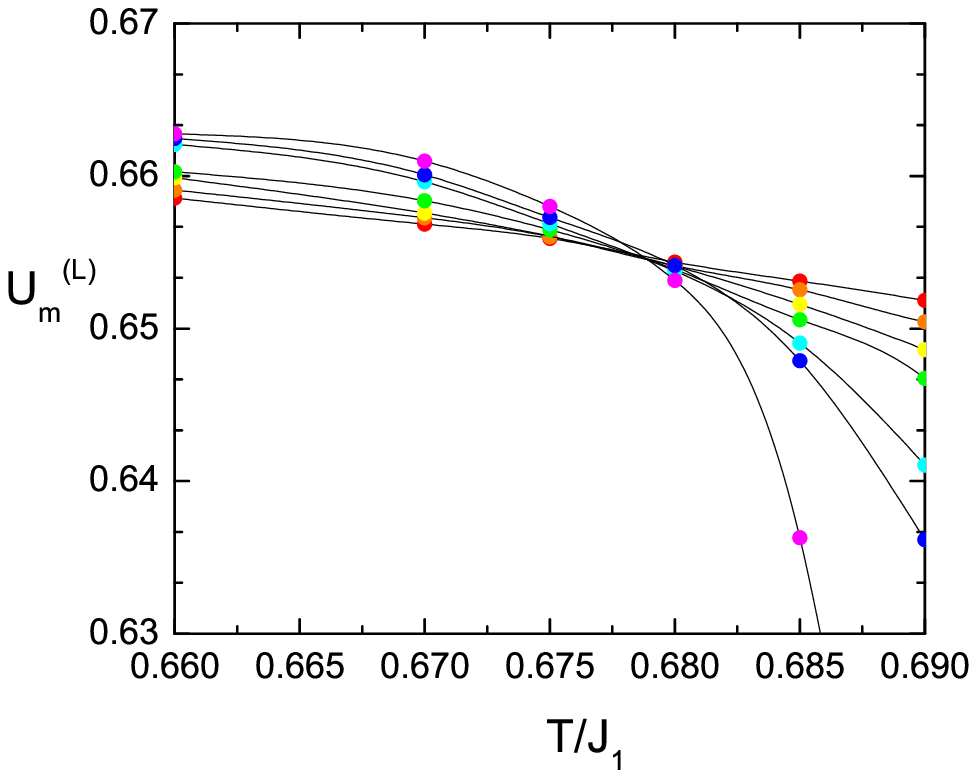}
    \vspace{-7mm}
    \caption{{\small Binder's cumulant $U_m(L)$ as a function of
    temperature, for $L=24,\ldots,90$.}}
    \label{fig6}}
\end{figure}

Exponents $\beta_k$ and $\gamma_k$ are found from scaling
properties of the order parameter $\overline{k}$ and the susceptibility
$\chi_k$ at the critical point
\begin{equation}
    \left(\bar{k}^{(L)}\right)_{T=T_I}\sim L^{\beta_k/\nu_k},\quad
    \left(\chi_k^{(L)}\right)_{T=T_I}\sim L^{\gamma_k/\nu_k},
\end{equation}
with the following result (see Fig.~\ref{fig5}):
\begin{equation}
    \beta_k=0.118(8), \quad \gamma_k=1.70(6).
\end{equation}
These values coincide within computational errors with exact
values of $1/8$ and $7/4$, correspondingly, of the 2D Ising model.
Using the scaling relations we find other exponents
\begin{equation}
    \alpha=2-2\nu_k=0.06(8),\quad
    \eta_k=2-\gamma_k/\nu_k=0.25(5).
\end{equation}
Note that the scaling relation
$\alpha+2\beta_k+\gamma_k=2.00(8)\approx2$ is satisfied within the computational error.

We have found also the universal value of the Binder cumulant
$U^*=0.615(6)$ at the critical temperature (see Fig.~\ref{fig2})
that is in agreement with the value $U^*\approx0.611$ observed in
the 2D Ising model \cite{2DIsing} with periodic boundary
conditions.

\subsection{BKT transition}
\label{bkt}

According to the Mermin-Wagner theorem \cite{Mermin} there is no
spontaneous magnetization at finite temperature in 2D magnets with
short range interactions and a continuous symmetry. But a
quasi-long-range order appears at non-zero temperature due to the
Berezinskii-Kosterlitz-Thouless mechanism \cite{BKT} in XY magnets
with $SO(2)$ symmetry.

It is important that as long as we measure the temperature in
units of $J_1$, the universal value of the jump $2T_{BKT}/\pi$
perturbs by the factor of $J_1/J_{\mathrm{eff}}$, where
$J_{\mathrm{eff}}$ is an effective coupling constant. The
competing exchange coupling $J_2$ gives rise to this factor. To
obtain it we considered the Coloumb-gas representation of the
model (\ref{ham}) using standard duality transformations
\cite{BKT, Duality}. As a result we obtain
\begin{equation}
    J_{\mathrm{eff}}=\sqrt{J_b\left(J_1-4J_2\right)}
        \label{jeffc}
\end{equation}
for the antiferromagnetic phase ($J_2<J_1/4$), and
\begin{equation}
    J_{\mathrm{eff}}=\sqrt{\frac12J_b\left(4J_2-\frac{J_1^2}{4J_2}\right)}
\end{equation}
for the helimagnetic one ($J_2>J_1/4$). Eq.~\eqref{jeffc} is in
accordance with results of Refs.~\cite{Garel, Kaplan}. In
particular, $J_{\mathrm{eff}}=\sqrt{3}/2$ for $J_2=0.5$.

A few authors have investigated the properties of the Binder
cumulant for magnetization as an alternative method of the BKT
transition temperature estimation \cite{Loison1}. Due to
finite-size corrections this method gives a value for the
transition temperature $T_{B}$ that is slightly larger than the
true value $T_{BKT}$. Nevertheless this method is useful as it
provides an estimation of the BKT transition temperature and an
extra evidence of separated transitions (if one finds that
$T_{B}<T_{I}$). Using the Binder cumulant crossing method
described above we find $T_{B}=0.679(2)$ (see Figs.~\ref{fig1} and
\ref{fig6}) that is $1.8\%$ smaller than $T_{I}$ given by
Eq.~\eqref{ti}.

\begin{figure}[t]
    \vspace{-7mm}
    \parbox{0.48\textwidth}{
    \centering
    \includegraphics[height=65mm]{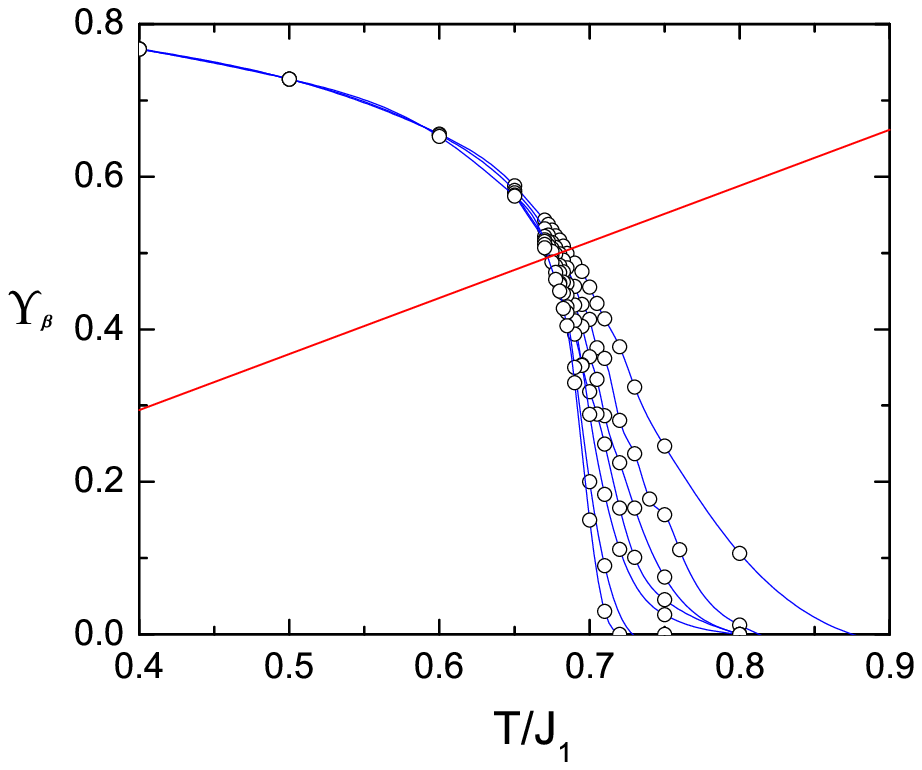}
    \vspace{-7mm}
    \caption{{\small Helicity modulus $\Upsilon_b$ in the $\mathbf{b}$ direction and its
    intersection with the line $2T/\pi J_{\mathrm{eff}}$.}}
    \label{fig7}}
    \hspace{0.02\textwidth}
    \parbox{0.48\textwidth}{
    \center
    \includegraphics[height=60mm]{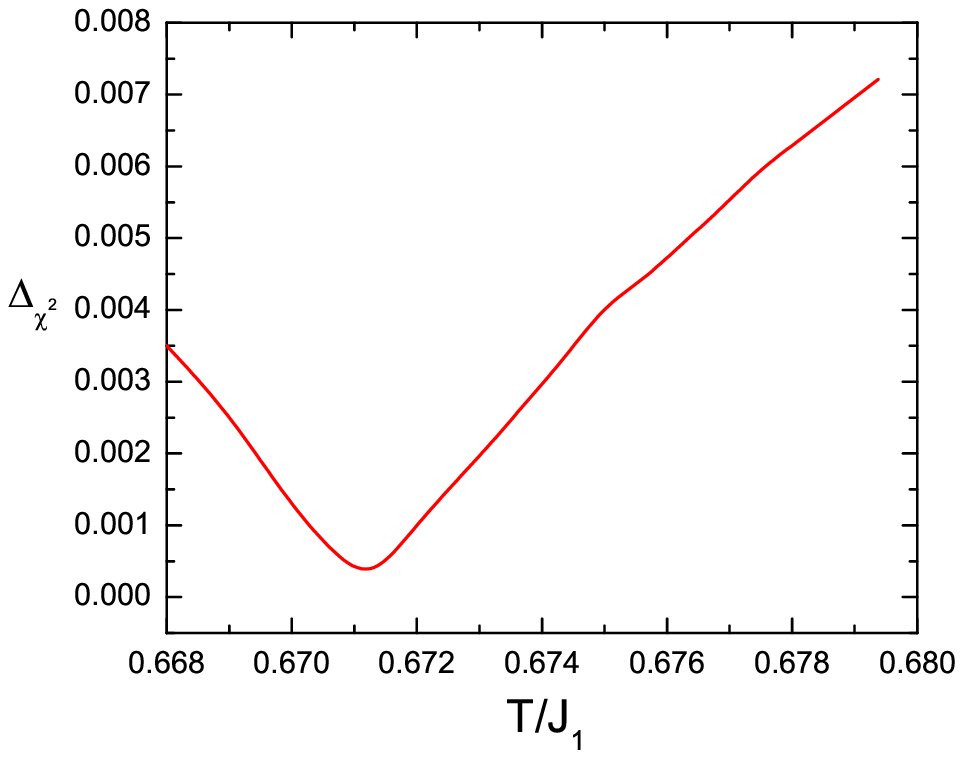}
    \vspace{-9mm}
    \caption{{\small Root-mean-square fit error $\Delta_c$ of the helicity
    modulus $\Upsilon_b$ to the Weber-Minnhagen scaling equation \eqref{Y=1+lnC}.}}
    \label{fig8}}
\end{figure}

\begin{figure}
    \vspace{-7mm}
    \parbox{0.48\textwidth}{
    \centering
    \includegraphics[height=65mm]{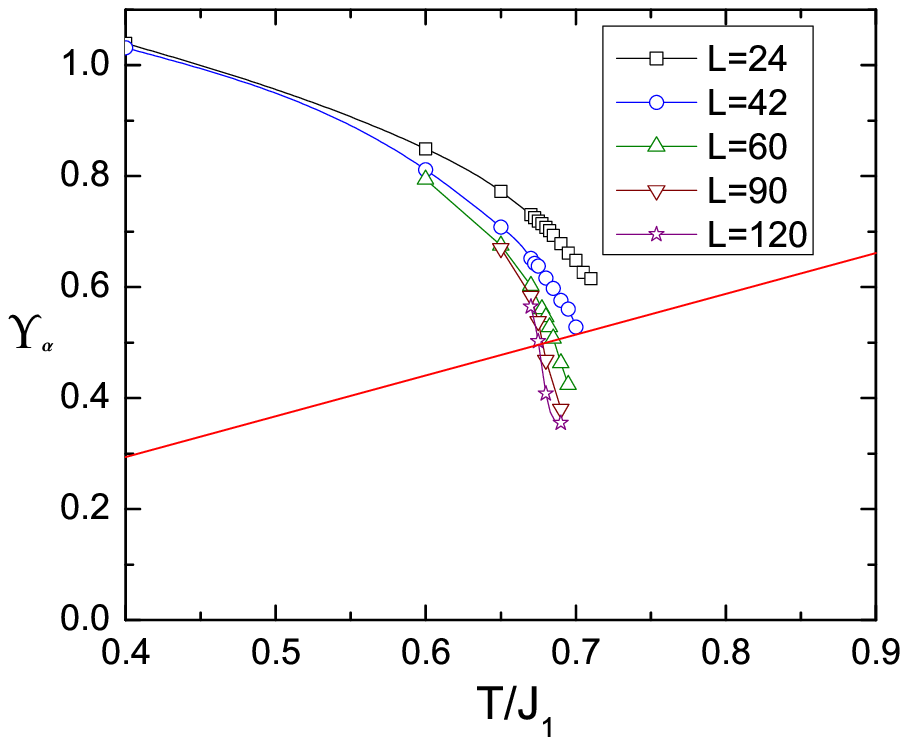}
    \vspace{-7mm}
    \caption{{\small Same as in Fig.~\ref{fig7} for $\Upsilon_a$.}}
    \label{fig14}}
    \hspace{0.02\textwidth}
    \vspace{-7mm}
    \parbox{0.48\textwidth}{
    \centering
    \includegraphics[height=60mm]{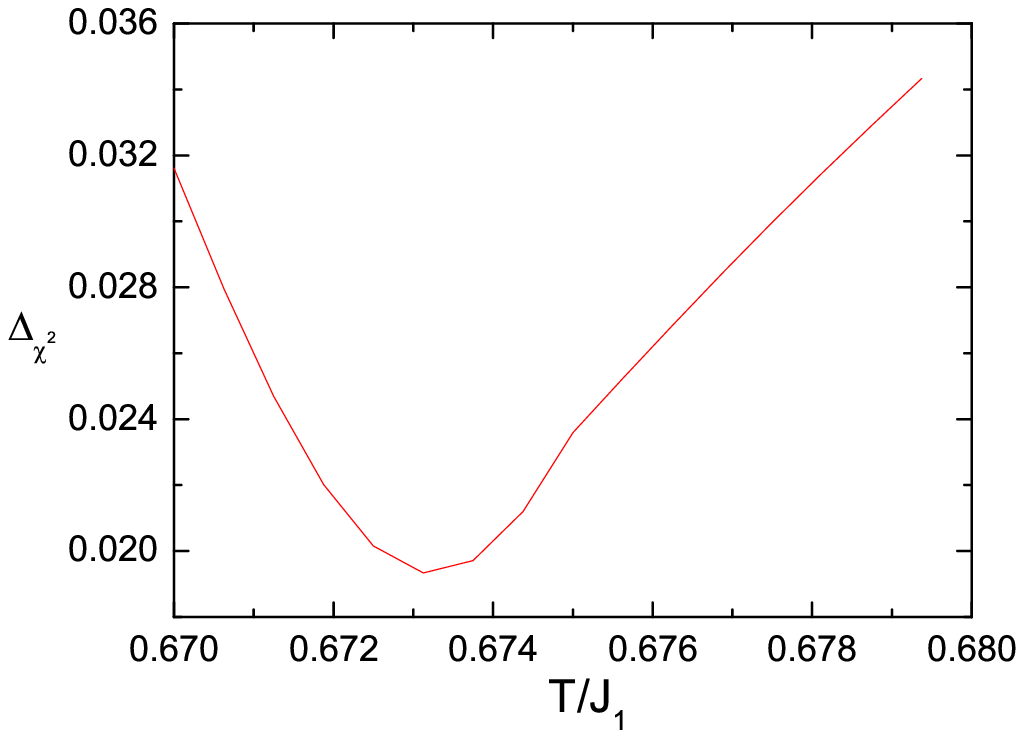}
    \vspace{-7mm}
    \caption{{\small Same as in Fig.~\ref{fig8} for $\Upsilon_a$.}}
    \label{fig142}}
    \vspace{5mm}
\end{figure}

To obtain $T_{BKT}$ precisely we use the Weber-Minnhagen
finite-size-scaling analysis \cite{weber} that is based on
consideration of logarithmic corrections to the value of the
helicity modulus at temperature close to $T_{BKT}$ having the form
\begin{equation}
    \Upsilon(T,L)=\frac{2T}{\pi J_{\mathrm{eff}}} \left(1+\frac1{2\ln L+c}\right),
    \label{Y=1+lnC}
\end{equation}
where $c$ is a fitting parameter. Fixing $T$ we find the
root-mean-square error $\Delta_c$ of the least-square fit of our
numerical data for $\Upsilon(T,L)$ with different $L\leq60$ that
is based on Eq.~(\ref{Y=1+lnC}). The minimum of $\Delta_c$
as a function of $T$ gives the value of the transition temperature
\cite{weber}. We obtain for the helicity modulus in the $\bf b$
direction (see Figs.~\ref{fig7} and \ref{fig8})
\begin{equation}
    T_{BKT}^{(\Upsilon_b)}=0.671(1).
        \label{tbktb}
\end{equation}

Corresponding results for $\Upsilon_a$ are shown in Figs.~\ref{fig14} and \ref{fig142}. For the
largest lattice of $L\geq48$ the Weber-Minnhagen finite-size scaling
analysis estimates the BKT transition temperature as
\begin{equation}
    T_{BKT}^{(\Upsilon_a)}=0.673(2)
\end{equation}
that is in agreement with Eq.~\eqref{tbktb}. It should be noted
that inaccuracy in estimation of $\Upsilon_a$ is larger than of
$\Upsilon_b$. That is why we use below the more precise value
\eqref{tbktb} for comparison between different methods.

\begin{figure}
    \parbox{0.48\textwidth}{
    \centering
    \includegraphics[height=65mm]{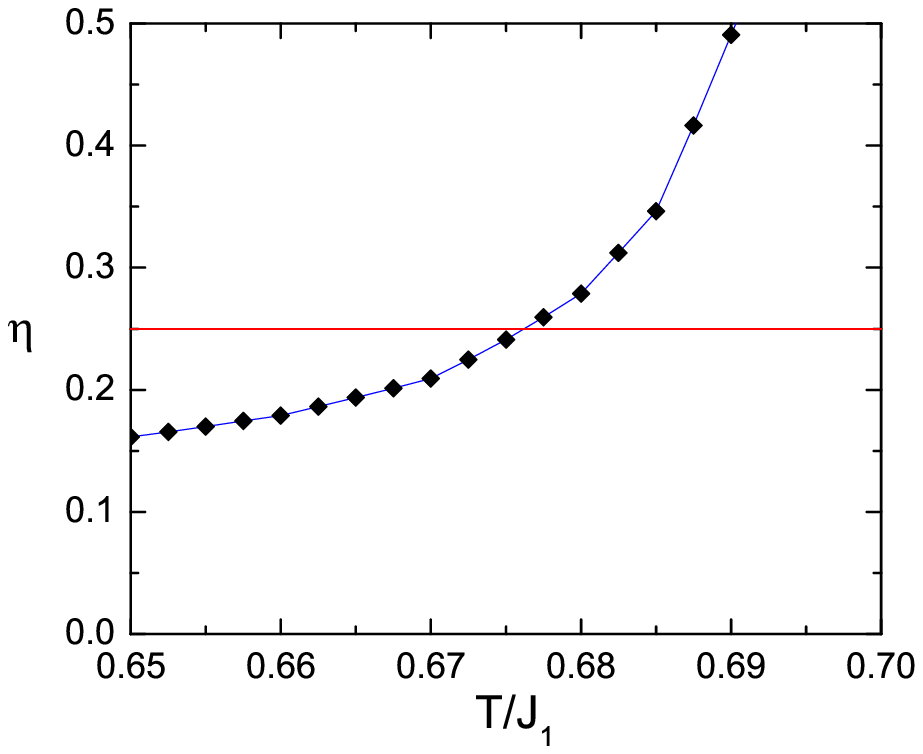}
    \vspace{-11mm}
    \caption{{\small Intersection of the exponent $\eta(T)$ with the bound $\eta=0.25$.}}
    \label{fig9}}
    \hspace{0.02\textwidth}
    \parbox{0.48\textwidth}{
    \centering
    \vspace{1mm}
    \includegraphics[height=60mm]{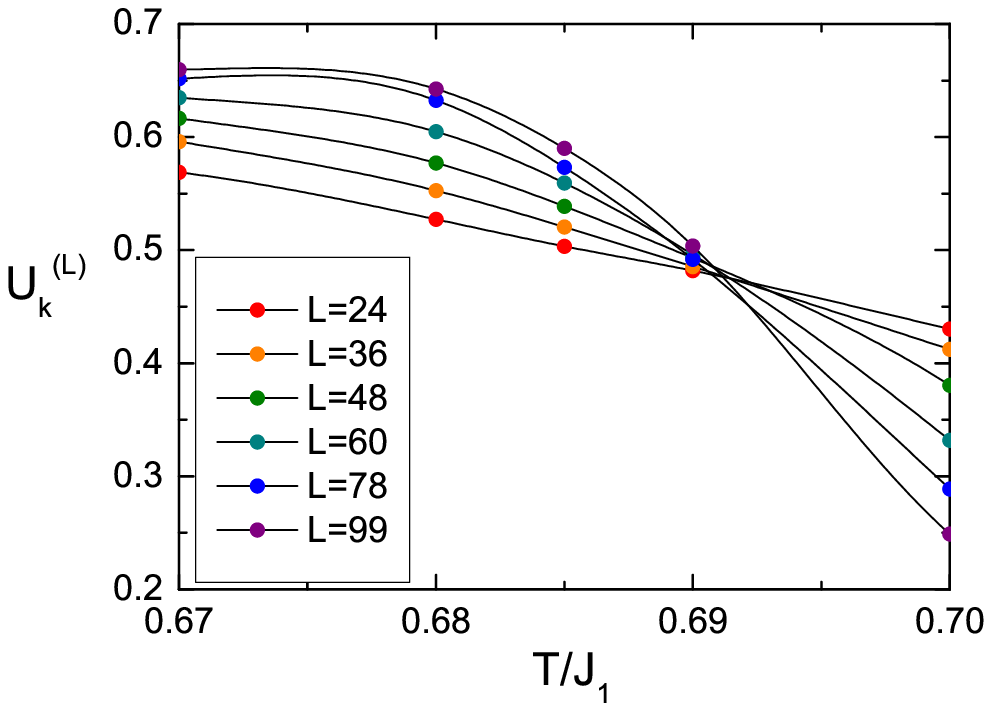}
    \vspace{-6mm}
    \caption{{\small Same as in Fig. \ref{fig2} for cylindric boundary conditions.}}
    \label{fig161}}
\end{figure}

To verify our results, we use also cylindric boundary conditions
with the periodic condition along the $\mathbf{b}$ axis and with
the free one along the $\mathbf{a}$ axis. We obtain results
consistent with those for periodic conditions. In particular,
transitions temperatures $T_{BKT}=0.671(2)$ and $T_I=0.6907(6)$
were estimated by the Weber-Minnhagen analysis and the Binder
cumulant crossing method (see Fig.~\ref{fig161}). The universal
value of the Binder cumulant  at the critical temperature is
$U^*=0.496(7)$ that is close to the value expected for the 2D Ising
model \cite{2DIsing} with mixed (cylindric) boundary conditions.

Another indication of the BKT transition is the equality to $0.25$
of the exponent $\eta(T)$. Below the transition temperature the
susceptibility diverges with the size of the system as
\begin{equation}
\chi_m(T,L)\sim L^{2-\eta(T)}.
\label{ch}
\end{equation}
The exponent $\eta$ as a function of temperature found using
Eq.~\eqref{ch} is shown in Fig.~\ref{fig9}. The intersection of
$\eta(T)$ with the bound $\eta=0.25$ gives
\begin{equation}
    T_{BKT}^{(\eta)}=0.676(2).
        \label{tbkte}
\end{equation}

Comparing Eqs.~\eqref{tbktb} and \eqref{tbkte} one notes that
$T_{BKT}^{(\eta)}>T_{BKT}^{(\Upsilon)}$. Because \cite{Jump}
$\eta(T_{BKT})=$ $T_{BKT} J_\mathrm{eff}/ 2\pi J_1
\Upsilon(T_{BKT})$,  we can not exclude that at the
true transition temperature the exponent $\eta$ and the jump of
the helicity modulus have non-universal values. Such a possibility
has been considered for other models from the class
$\mathbb{Z}_2\otimes SO(2)$. \cite{Teitel, FFXY, Triangular, Gas,
nonuniv} If it is so $\eta$ has a value smaller than 0.25 and the
jump is greater than $2J_1/\pi J_\mathrm{eff}$. Thus, our data
show that $\eta(T_{BKT}^{(\Upsilon)})\approx0.22$.

\subsection{Neighborhood of the Lifshitz point and the phase diagram}
\label{phsec}

Besides the case of $J_2=0.5$ considered above in detail we have
carried out similar discussions of $J_2=0$, 0.1, 0.309 and 1.76 to
obtain the phase diagram shown in Fig.~\ref{phase}. Some results
of this consideration are summarized in Table~\ref{table}. The
case of $J_2=0$ corresponds to the well known XY model on a square
lattice and we find $T_{BKT}=0.891(2)$ that is consistent with the
previous results \cite{MCforKT}.

\begin{figure}[t]
    \centering
    \includegraphics[height=90mm]{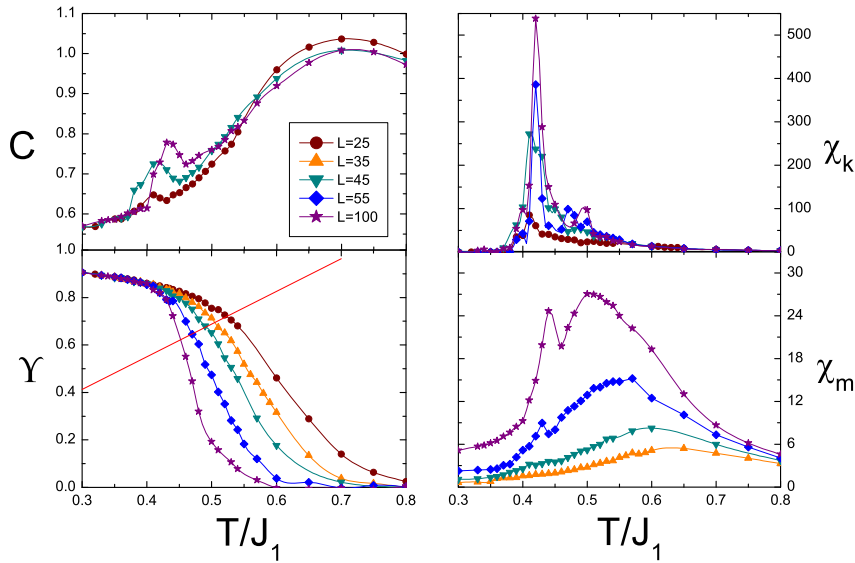}
    \vspace{-7mm}
    \caption{{\small Specific heat $C(T)$, helicity modulus $\Upsilon_b$ and susceptibilities $\chi_{k,m}$ for $J_2\approx 0.309$.}}
    \label{fig163}
\end{figure}

It should be stressed that
we obtain $T_I>T_{BKT}$ at $J_2>1/4$ in accordance with
conclusions of Refs.~\cite{Korshunov,Kolezhuk} and in contrast to
Refs.~\cite{Garel,Cinti}. Because this our finding is at odds with
that of the similar numerical consideration of the same model
carried out in Ref.~\cite{Cinti}, our special interest is to
consider the case of $J_2\approx 0.309$ that is close to $J_2=0.3$
discussed in Ref.~\cite{Cinti}.

\begin{table}
\caption{Some results of our discussion of the model \eqref{ham}. Here $L_{max}$ is the maximum value of $L$ considered.}
\centering
\begin{tabular}{ccccc}
\hline
\hline
$J_2/J_1$ & $\theta_0$ & $L_{max}$ & $T_{BKT}/J_1$ & $T_I/J_1$\\
\hline
$0$&  0  &40& 0.891(2) & - \\
$0.1$& 0 &30& 0.781(3) & - \\
$\approx0.309$& $4\pi/5$ &100& 0.443(5) & 0.48(1)\\
0.5& $2\pi/3$ &150& 0.671(1) & 0.690(1)\\
$\approx1.76$ & $6\pi/11$ &66& 1.24(1) & 1.285(7)\\
\hline
\hline
\end{tabular}

\label{table}
\end{table}

The case of $J_2\approx 0.309$ corresponds at $T=0$ to
$\theta_0=4\pi/5$. In particular, we find $T_{BKT}=0.443(5)$ that
is very close to the value reported in Ref.~\cite{Cinti}.
Estimating the temperature of the Ising critical point by the
Binder cumulant crossing method, we encounter the anomalous
behavior of the chiral order parameter and do not obtain a
reliable result. Apparently, it is the reason why the authors of
Ref.~\cite{Cinti} base their conclusion about the Ising transition
on the behavior of the specific heat $C(T)$ and susceptibilities
$\chi_{k,m}$. Our results for these quantities are shown in
Fig.~\ref{fig163} and they are consistent with those of
Ref.~\cite{Cinti}. It is seen from Fig. \ref{fig163} that the
chiral susceptibility has a high peak at $T\approx 0.4$, while the
specific heat and the magnetic susceptibility have subsidiary
peaks at $T\approx 0.4$ which grow with the lattice size
increasing. These anomalies at $T\approx 0.4<T_{BKT}$ are
attributed in Ref.~\cite{Cinti} to the Ising transition.

However, we observe that for other $J_2>1/4$ the specific heat has
only one peak corresponding to the logarithmic divergence which
characterizes the Ising transition (see, e.g., Fig.~\ref{fig3} for
$J_2=0.5$). Therefore, the behavior of $C(T)$ shown in
Fig.~\ref{fig163} is not normal for the model discussed and it is
characteristic of the Lifshitz point neighborhood. To account for
this anomaly we examine the behavior of the chiral order parameter
in detail.

\begin{figure}[t]
    \vspace{-7mm}
    \parbox{0.48\textwidth}{
    \center
    \includegraphics[height=65mm]{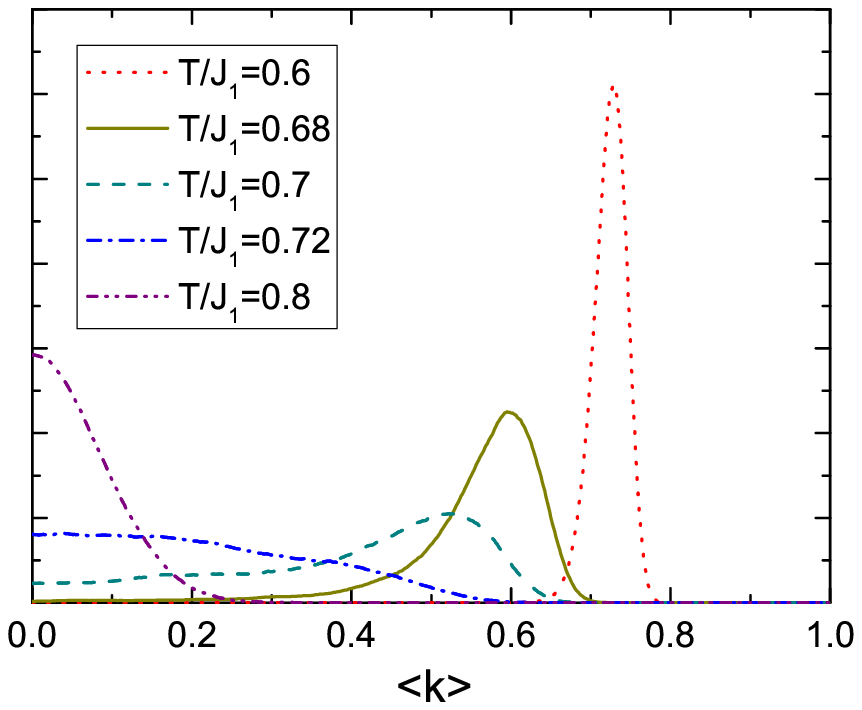}
    \vspace{-7mm}
    \caption{{\small The chiral order parameter distribution for $J_2=0.5$ and $L=42$.}}
    \label{fig17}}
    \hspace{0.02\textwidth}
    \parbox{0.48\textwidth}{
    \centering
    \includegraphics[height=65mm]{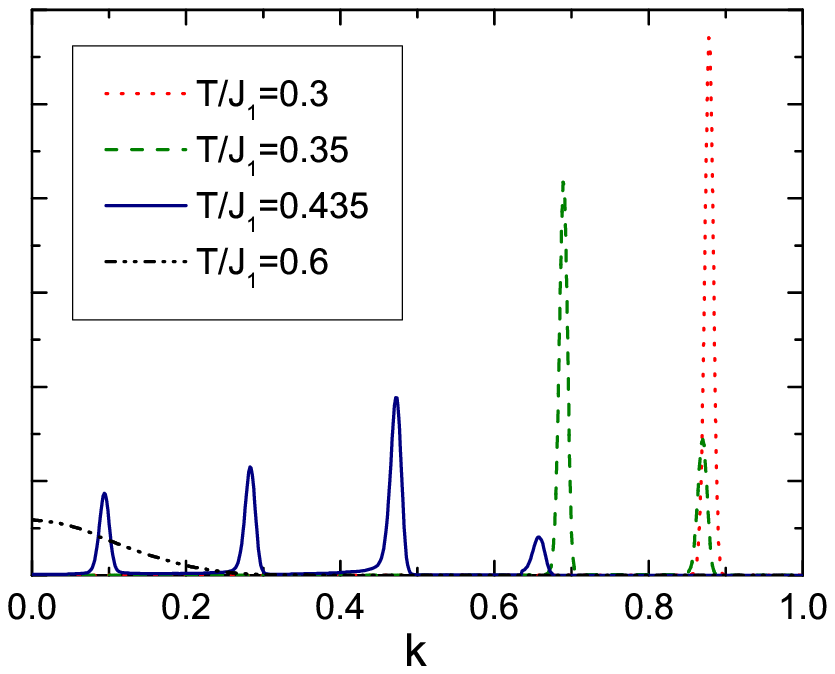}
    \vspace{-7mm}
    \caption{{\small The chiral order parameter distribution for $J_2=0.309$, $L=45$ at different
    temperatures.}}
    \label{fig18}}
\end{figure}

Fig. \ref{fig17} shows the chiral order parameter distribution for
$J_2=0.5$ and $L=45$. It looks like a customary order parameter
distribution in a system with the second order transition. In
particular, the distribution has a Gaussian form below a critical
temperature with a peak at $\bar{k}$ (see the curve for $T=0.6$).
In approaching to the critical temperature the distribution
acquires an appreciable tail (see curves for $T=0.68$ and
$T=0.7$). Such a broad distribution leads to a peak in the
susceptibility. The distribution has a peak at $k=0$ above the
critical point (see curves for $T=0.72$ and $T=0.8$).

\begin{figure}[t]
    \vspace{-7mm}
    \parbox{0.48\textwidth}{
    \centering
    \includegraphics[height=65mm]{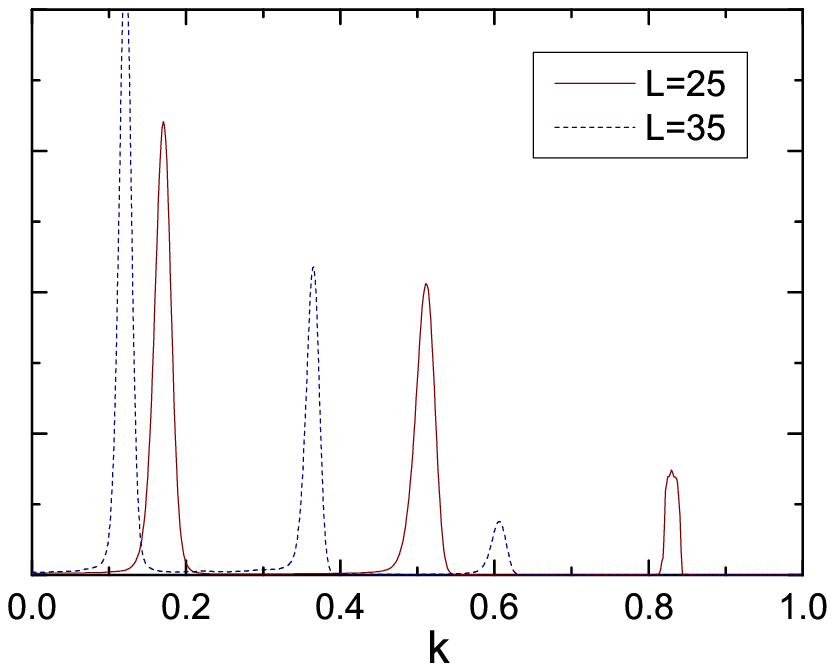}
    \vspace{-7mm}
    \caption{{\small The chiral order parameter distribution for $J_2=0.309$, $L=25$ and $L=35$
    at $T/J_1=0.43$.}}
    \label{fig20}}
    \hspace{0.02\textwidth}
    \parbox{0.48\textwidth}{
    \center
    \includegraphics[height=65mm]{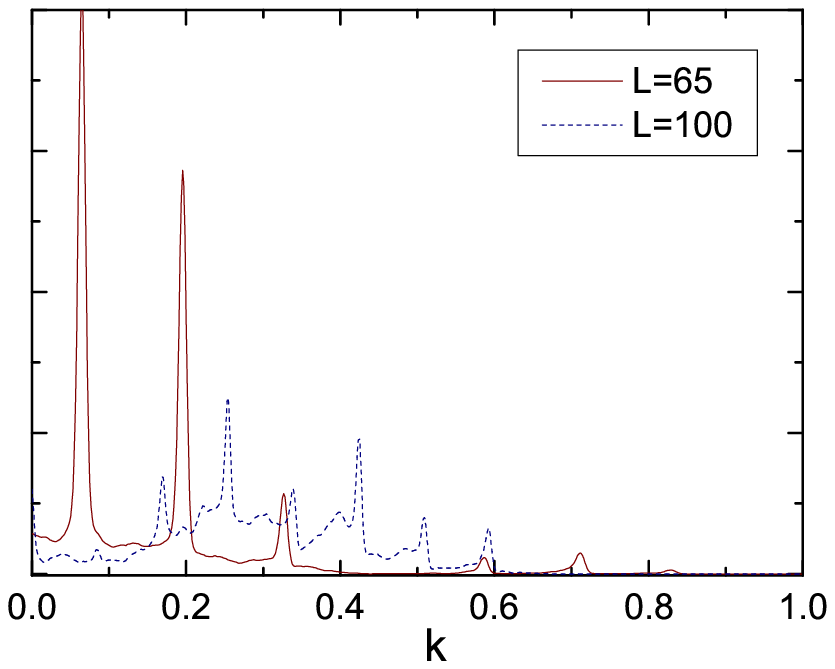}
    \vspace{-7mm}
    \caption{{\small The chiral order parameter distribution for $J_2=0.309$, $L=65$ and $L=100$
    at $T/J_1=0.43$.}}
    \label{fig21}}
\end{figure}
\begin{figure}[t]
    \vspace{-7mm}
    \parbox{0.48\textwidth}{
    \centering
    \includegraphics[height=65mm]{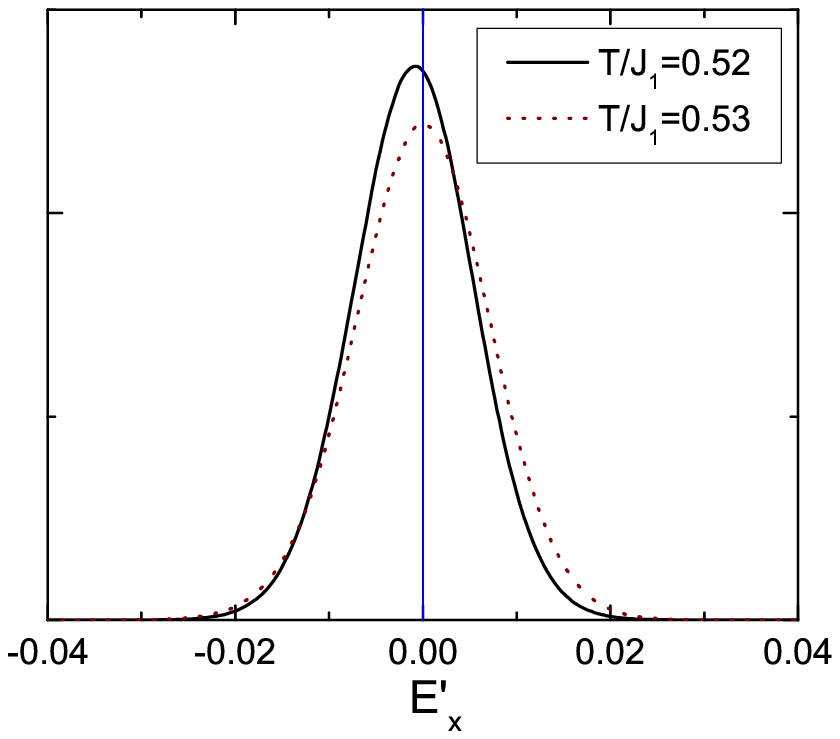}
    \vspace{-7mm}
    \caption{{\small Distribution of $E_a'$ defined in Eq.~\eqref{Yx} for $L=65$.}}
    \label{fig22}}
    \hspace{0.02\textwidth}
    \parbox{0.48\textwidth}{
    \center
    \includegraphics[height=60mm]{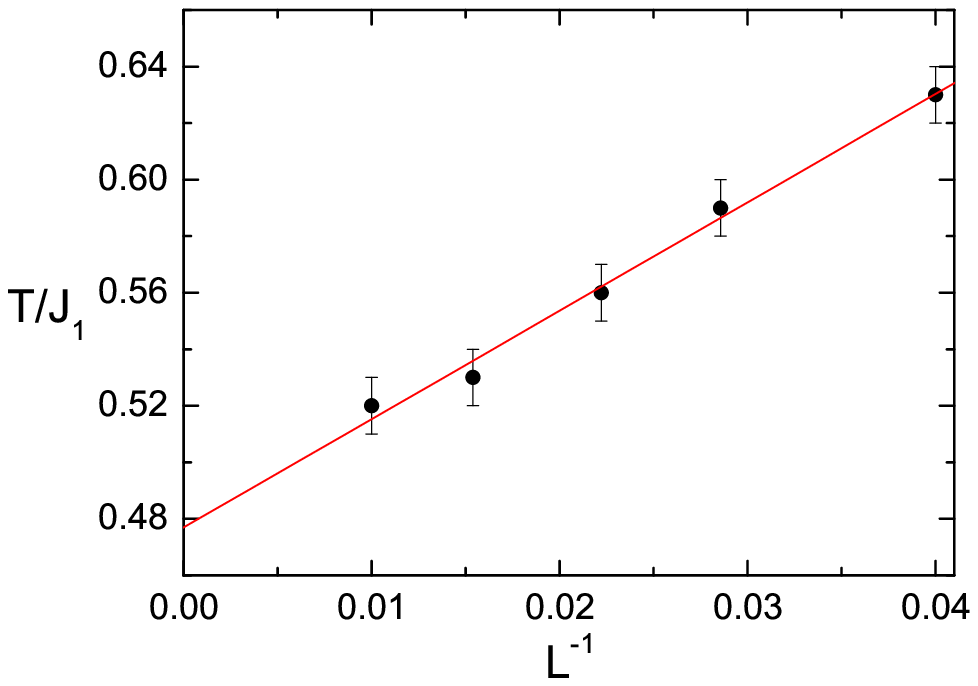}
    \vspace{-7mm}
    \caption{{\small Estimation of $T_I$ for $J_2=0.309$ that is based on Eq.~\eqref{extr}.}}
    \label{fig23}}
\end{figure}

However the picture for the chiral order parameter distribution is
quite different close to the Lifshitz point, e.g., at $J_2\approx
0.309$. We show in Fig.~\ref{fig18} the chiral order parameter
distribution for $L=45$ and different temperatures. One can see
one Gaussian peak at $T<0.35$ in agreement with the common picture
described above. But a few additional peaks arise at $T>0.35$.
Then, the number and the breadth of peaks depend on the lattice
size $L$ (see Figs.~\ref{fig20} and \ref{fig21}) and, what is much
more important, on the boundary conditions. For the cylindrical
boundary conditions, these peaks are broader and they are
accompanied by great number of accessory peaks. Such a
distribution of the chiral order parameter is characteristic in
the case of $L=45$ to the range of temperature from $T\approx0.35$
to $T\approx0.55$. One can see from Fig.~\ref{fig18} that at
$T=0.57$ the distribution has a form that is typical for a
disordered phase with the peak at zero value of the order
parameter. Then, it is clear that the order parameter distribution
at $T\approx0.4$ shown in Fig.~\ref{fig18} does not correspond to
a critical point of a continuous transition.

Apparently, the origin of such a peculiar behavior at $J_2\approx
0.309$ are metastable states with different values of the chiral
order parameter which prevent the system investigation
considerably. The multi-peak structure of the order parameter
distribution leads to a sudden jump of the susceptibility and
states intermediate between metastable configurations give rise to
the specific heat anomaly. It should be noted that energy values
of the metastable states are close since we observe in our
simulations that the energy distribution has a Gaussian form even
for the largest lattice size.

To estimate the Ising transition temperature at $J_2\approx 0.309$
we analyze $E'_a$ distribution defined in Eq.~\eqref{Yx} and
discussed in Sec.~\ref{helmod}. As it is pointed out above,
$\langle E'_a\rangle\ne0$ whenever a helical ordering exists.
Fig.~\ref{fig22} shows that the distribution is not symmetric
relative to zero at $T=0.52$ and $L=65$, while it is definitely
symmetric at $T\ge0.53$. Therefore, the critical temperature for
$L=65$ can be roughly estimated as $T_I(L)=0.53(1)$. The
transition temperature can be estimated using the relation
\begin{equation}
\label{extr}
T_I=T_I(L)-\frac{A}{L^\nu},
\end{equation}
where $A$ is constant and $\nu=1$ as it is expected for an Ising
transition. An extrapolation to the thermodynamic limit using
Eq.~\eqref{extr} gives $T_I=0.477(12)$ (see Fig.~\ref{fig23}).
Then, we obtain that $T_{BKT}<T_I$ even in the neighborhood of the
Lifshitz point.

\section{Conclusion}
\label{conc}

We discuss critical properties of the 2D helimagnet described by
the Hamiltonian \eqref{ham}. It belongs at $J_2/J_1>1/4$ to the
same class universality as the fully frustrated XY model and the
antiferromagnet on triangular lattice which have two successive
phase transitions upon the temperature decreasing: the first one
is associated with breaking of the discrete $\mathbb{Z}_2$
symmetry and the second one is of the BKT type at which the
$SO(2)$ symmetry breaks. We confirm that this scenario is realized
also in the model \eqref{ham} at $J_2/J_1>1/4$ and obtain the
phase diagram shown in Fig.~\ref{phase}. A narrow region exists on
this phase diagram between lines of the Ising and the BKT
transitions that corresponds to the chiral spin liquid.

In particular, we demonstrate that the number and sequence of
transitions do not depend on the turn angle $\theta_0$ of the
helix twist at $T=0$. Then, this quantity is not a critical
parameter that has been already found \cite{Helix, Kawamura2} in
three-dimensional helimagnets. We find that it is useful in
numerical discussion of helimagnets to choose parameters of the
Hamiltonian so that the helix pitch at $T=0$ to be commensurate.
It allows to use definition \eqref{m} of the magnetic order
parameter in which the summation over sublattices is involved.

We find in accordance with results of Ref.~\cite{Cinti} that the
specific heat and susceptibilities have subsidiary peaks at low
$T$ near the Lifshitz point $J_2/J_1=1/4$ (see Fig.~\ref{fig163}).
However, in contrast to the conclusion of Ref.~\cite{Cinti} we
demonstrate that these anomalies do not signify a continuous phase
transition. Apparently, their origin is in metastable states near
the Lifshitz point which lead also to a peculiar distribution of
the chiral order parameter shown in Fig.~\ref{fig18}.\bigskip

This work was supported by the RF President (grant MD-274.2012.2),
the RFBR grant 12-02-01234, and the Program "Neutron Research of
Solids".

\end{document}